# ANISOTROPIC REINFORCEMENT OF NANOCOMPOSITES TUNED BY MAGNETIC ORIENTATION OF THE FILLER NETWORK

Jacques Jestin[1], Fabrice Cousin[1], Isabelle Dubois[1], Christine Ménager[2], Ralf Schweins[3], Julian Oberdisse[4], François Boué[1].

[1] Laboratoire Léon Brillouin (LLB), CEA/CNRS, CEA Saclay 91191 Gif-sur-Yvette France.

[2] Laboratoire Liquides Ioniques et Interfaces Chargées (LI2C), UMR 7612 CNRS / Université Paris 6, 4, place Jussieu - case 63 75252 Paris cedex 05 France.

[3] Institut Laue-Langevin, DS/LSS, 6 rue Jules Horowitz, B.P. 156, 38042 Grenoble Cedex 9 France.

[4] Laboratoire des Colloïdes, Verres et Nanomatériaux (LCVN), UMR 5587, CC 26 Université Montpellier II, 34095 Montpellier Cedex 05 France.

**Abstract**

We present a new material which displays anisotropic and mechanical properties tuneable during synthesis under magnetic field. It is formulated by mixing aqueous suspensions of polymer nanolatex and magnetic nanoparticles, coated by a thin silica layer to improve their compatibility with the polymeric matrix, followed by casting. The magnetic properties of these nanoparticles enable their pre-orientation in the resulting nanocomposite when cast under magnetic field. Detailed insight on dispersion by Small Angle Neutron Scattering (SANS) shows chainlike nanoparticle aggregates aligned by the field on the nanometer scale. Applying strain to the nanocomposite parallel to the particle chains shows higher mechanical reinforcement, than when strain is transverse to field. . SANS from strained samples shows that strain parallel to the field induce an organization of the chains while strain perpendicular to the field destroys the chain field-induced ordering. Thus improved mechanical reinforcement is obtained from anisotropic interconnection of nanoparticle aggregates.

**Introduction**

Progress in both design of innovative materials and understanding of their properties has been achieved in the past two decades by extending polymer reinforcement by fillers to nano-fillers[1]. Among the still open challenges, the improvement and possibly tuning of a specific material property through a simple external trigger is of particular importance. A promising pathway is to bring in a new degree of freedom, e.g., by using magnetic, instead of classical inorganic particles as filler. The design of a functional nanocomposite material with the possibility to orient the filler inside the matrix should allow to tune macroscopic material properties, and in particular the mechanical ones, using a simple external magnetic field. This concept has been explored in the past twenty years by many research teams, mostly studying the effect of micron-size particles or fibres (needles)[2-5]. Two routes are possible: applying the field on the material after synthesis or applying it during the synthesis. We use here this second approach for nanoparticles; in such case, the question of the mechanical consequences of local anisotropy due to specific orientation of a nanofiller, i.e. particles with a size comparable to polymer network mesh sizes, has still to be solved.

Here we present a new material based on a dispersion of magnetic nano-particles in a polymer matrix. This is achieved in a controlled way by adapting the knowledge on these systems in solution ("Ferrofluids")[6] to our own method of dispersion of silica nano-particles in a polymer matrix made of latex beads[7-9]: via solvent (water) casting of a suspension of magnetic nanoparticles and polymer latex, a "Ferrolatex nano-composite" is formed. In this article, we will show that achieving these steps under the action of a weak magnetic field induces orientation responsible of a strong anisotropy of the reinforcement. Moreover, this mechanical anisotropy can be understood by measuring the spatial arrangement and orientation of the particles on a scale between one nanometer and two hundred nanometers, using Small Angle Neutron Scattering (SANS).

**Sample preparation**

The synthesis of a composite made of a latex film filled with magnetic particles requires two main steps (fig. 1). The first main step is the synthesis of the magnetic particles according to a well known process described in the Methods Section. In a second time, in the same batch, the particles are covered with a thin layer of silica. This silica shell makes the particles pH compatible with the latex beads while keeping the initial size and the magnetic properties of the core. Hence we can use the technique developed for introducing silica particles in a latex film[7,9]. The colloidal suspension of particles is mixed with a polymer bead (R~200 Å) suspension at a fixed, high pH, followed by evaporation of the (aqueous) solvent. The latex nanospheres come in contact, and the capillary forces compress them into adjacent polyhedrons. We know[7] that under deformation above Tg, such matrix deforms affinely (homothetically) to the macroscopic deformation, due to the elasticity of the membrane.

The colloidal mixture, with a proportion of magnetic nano-particles to the latex spheres of 5% by volume is poured into rectangular $1\times6\times0.1$ cm$^3$ moulds and dried (at T ~ 60°C), i.e. above the Film Formation Temperature which is itself close to the glass temperature transition of the latex (Tg ~ 33°C).

During evaporation, a weak constant field (150 gauss) is applied in order to orient the magnetic filler. A similar sample is also evaporated in the same mould without magnetic field. We call it the "zero-field cast" sample.

**Mechanical reinforcement**

We have performed mechanical tests with a home-made uniaxial elongation machine. Keeping in mind the direction of the magnetic field during casting, we have the opportunity to study two simple but fundamental cases, stretching parallel or perpendicular to the field, using two rectangular films (1x6x40 mm$^3$) cut from the same sample. This allows a direct comparison of the stress-strain isotherms in the two directions. The samples are first immersed in a silicone oil bath at a temperature $T_g$ + 27°C, then, elongated at constant velocity gradient. When the desired elongation is reached, namely $\lambda$=1.5, 2 and 3, the samples are immediately frozen by immersion in a carbonic ice bath. The reference case for the mechanical tests is given by the stress-strain isotherm of the pure latex matrix. It is used as a normalisation, i.e. we divide by all the stress-strain isotherms measured for the nano-composite: this gives us a reinforcement factor R ($\lambda$) for each composite with respect to its own matrix. To check the reproducibility of the synthesis of the magnetic nanoparticles, we have pulled a second series of films synthesized out of a second batch of magnetic nano-particles with similar sizes. The four stretching curves $\sigma$ versus $\lambda$ are plotted on Fig. 2a together with the ones of the pure matrix and of the zero-field cast film. For all samples with nanoparticles, the shape of the curves is similar: compared to the pure latex, they show a fast increase at the beginning of the elongation, while the shape at large $\lambda$ tends to be similar. There is a very strong effect for the zero-field cast sample, whereas after field-casting we can observe that the reinforcement sample is lower by one order of magnitude.

But the striking effect is the difference between the field-cast samples stretched in parallel and perpendicularly to the magnetic field. They show a very different behaviour, especially at low deformation. Moreover, the mechanical tests for the two batches of magnetic nano-particles show a very nice reproducibility for the two directions.

If we plot in Fig. 2b the reinforcement factor R ($\lambda$) as function of the deformation $\lambda$, we see a strong reinforcement for the small deformation and a rapid decrease of R ($\lambda$) with $\lambda$ toward a constant value, noticeably smaller (but larger than one). This is very similar to what formerly described for nanolatex composites filled with silica nanoparticles[9]. The curve aspect is common to the two directions. The important fact is that **the reinforcement factors obtained are much higher when the field is along the deformation direction than when the field is perpendicular to it.** We quantify the anisotropy of the mechanical properties of a given sample by the ratio of the parallel reinforcement to the perpendicular reinforcement (Fig. 2c). **It is more than a factor two**. It has also to be noted for the discussion below that such anisotropy in R ($\lambda$) vanishes for large elongations ($\lambda > 2$).

In summary, we are able to obtain a strong effect of the mechanical properties of the nanocomposites under orientation by the field. The question is now to try to correlate the macroscopic effect with the local organization of the filler network in the matrix and its evolution under the mechanical strain. This is reported in the following using SANS measurements analysis.

**Local structure**

SANS measurements have been performed on samples after each successive step of the synthesis. In the following, figures from 3 to 5 will be presented in the same way: above 2D intensity maps, in the middle the radially averaged intensities as a function of the scattering vector, and at the bottom a sketch in the real space illustrating the structure of the network of aggregates. First, Fig. 3a shows results for magnetic particles in dilute aqueous solution at low volume fraction ($\Phi$ = 0.001) before and after the silica-covering process. The scattering is obviously isotropic and its radial-average can be fitted at large q to the one of spherical primary particles with a log-normal polydisperse distribution of average radius 35 Å and variance $\sigma$ = 0.3. The low q rod-like scattering ($q^{-D_f}$, fractal dimension $D_f$ = 1) suggests that they are organized in small "**primary clusters**". The mean number of particles $N_{prim}$ in a cluster is of the order of 3 to 4 particles, as obtained from the fit to the scattering of a fractal aggregate of $D_f$ = 1 (cf. ref.[10] for details). Second, patterns after casting and before stretching, are presented in Fig. 3b. As expected, the zero-field cast sample is isotropic. The radial average displays the characteristic signature of the formation of aggregates during film drying. Fortunately, the size of such large objects is observable in our q-window, as illustrated by a plateau in the scattering curve at low q. This intensity limit corresponds to an aggregation number of around 10 primary clusters (~ 30 nanoparticles). At slightly larger q, where the internal structure of the aggregates is probed, we observe a $q^{-2}$ behaviour, i.e. a fractal dimension $D_f$ ~ 2, as typically encountered for either Diffusion Limited Aggregation[11] (DLA, $D_f$ = 1.8 in 3D) or Reaction Limited Aggregation[12] (RLA, $D_f$ = 2.1 in 3D). At larger values of q (around $4.10^{-2}$ Å$^{-1}$), we see an oscillation in the scattering curve which corresponds to the correlations between primary clusters inside the aggregates. Let us now come to the field-cast film: it gives a nicely anisotropic scattering pattern (Fig. 3c). The elliptic shape with large horizontal axis, i.e. transverse to the field, is characteristic of objects elongated vertically: thus there are aggregates aligned parallel to the field. To quantify this effect, let us radial average the scattering inside an angular sector **along one direction**, either **vertical** or **horizontal**, with a width of 25°. For small q, the intensity is smaller along the vertical axis than along the horizontal one. The aggregates are aligned on a larger scale along the field (vertical here) and we can assume that they form "chains of aggregates" inside the matrix. The intensity along the horizontal axis shows a strong $q^{-3}$ decrease. This exponent 3 is mostly encountered for compact objects, composed here of chains of aggregates that we will define as "supra-aggregates". The scattering is still increasing at the lowest q of the experimental window. This means that the size of the supra-aggregates is too large to be probed here. The sizes of the structural organization of the aggregates inside one supra-aggregate are also too large to be accessible within the experimental q-range. For the high-q values, the intensities along the vertical and the horizontal axis are superimposed with the zero-field sample. This indicates that the local structure of the aggregates, composed of the primary clusters, is not affected by the application of the magnetic field. To summarise this section, we have succeeded in orienting the filler as chains of aggregates, inside the nanolatex composite using a constant field.

Let us now discuss the evolution of the structure of the filler network under stretching. We start with the first case, where to the elongation (vertical) is parallel to the field  Figure 4 shows the 2D intensity map for each λ. When the elongation rate increases, the anisotropy is increasing towards a surprising clear cut hexagonal form. This hexagonal form is due to the rise of a maximum in the vertical direction. This maximum appears at a small q value for the first elongation (λ = 1.5) and its position moves to higher q values at higher elongations. This behaviour is highlighted in the S(q).q² versus q representation of Fig. 4a and 4b, convenient here since the scattering decays like $q^{-2}$ at intermediate q. S(q) is the result of the division of the I(q) by the form factor P(q) of the primary clusters. The data show that elongation of the sample induces organization of the aggregates with a characteristic distance between the aggregates in the vertical direction. Please note that the curve for the unstretched sample showed no maximum in the available q range. Since the correlation peak is shifted towards higher q when the elongation increases, the characteristic distance decreases. Intuitively, one would expect the contrary, as the aggregates move away from each other along the stretching direction. This implies that under strong deformation some aggregates coming from neighbouring chainlike supra-aggregates intercalate vertically between two aggregates. Such a mechanism results in an effective decrease of the vertical inter-aggregate distance, as illustrated in the real space sketches of Figure 4. Along the horizontal direction, the intensity decreases compared to the isotropic signal: the supra-aggregates are brought closer due to the lateral shrinking of the sample which increases the effective compressibility between them.

In Fig. 5, we present the second case where the stretching direction (here horizontal) is perpendicular to the chainlike supra-aggregates. In addition to the isotropic intensity map, two elongation ratios are presented, λ=1.5 and λ=2. The result of the deformation is completely different in this case compared to the previous one; the scattering pattern becomes **isotropic** as the elongation is increased. Here again this is highlighted when one plots S(q).q² as function of q along both vertical and horizontal directions of the 2D intensity map: the anisotropy is important for the non-stretched sample (Fig. 5a) and becomes small for the elongated one (Fig. 5b). The orientation of filler breaks under the mechanical constraint: the aggregates are reoriented isotropically. Moreover, the supra-aggregates are destroyed: the final scattering of sample stretched to λ= 2 superimposes nicely to the scattering from the zero-field cast sample (black symbols).

Finally, when confronting the structural organization of the filler network deduced from SANS to the reinforcement, a consistent picture emerges. Under stretching, the parallel oriented chains rearrange along the deformation direction due to a contraction of the interconnecting particle network. In contrary, the perpendicularly oriented network breaks under deformation. This leads to a higher reinforcement factor in parallel direction. In the perpendicular direction, the final structure of the chains is resumed to the organization of the zero-field cast sample (at least in the available q range), but the latter has a higher reinforcement factor. We propose to attribute this difference to the existence of larger aggregates connected at larger scales.

**Conclusion**

To summarize, we have designed an original material which is structured on the nanoscale and which displays anisotropic and tuneable mechanical properties. The samples have been formulated rather easily by mixing aqueous suspensions of polymer and magnetic colloids, followed by casting. These nanocomposites bear magnetic properties, and therefore the possibility of pre-orientation when cast under field. Detailed insight on dispersion by Small Angle Neutron Scattering shows that primary particle aggregates are organized at the nanometer scale, and align in the presence of a field. Such anisotropy has a clear effect on the mechanical reinforcement brought by the fillers. The interesting novelty is the anisotropic reinforcement: it is larger by a factor 2 at low deformation along the field than transverse to it. Microscopically, this can be understood with the SANS-data: while strain perpendicular to the chains destroys chain ordering, strain parallel to the field induces additional organisation. Thus anisotropic mechanical reinforcement is obtained through anisotropic interconnection of nanoparticles, which is a one-dimensional version of the filler network effect. We note that all these effects vanish at large $\lambda$ (>2).

In addition to such controlled anisotropy of our nanocomposites, we would like to point out another contribution of this study, which concerns the conceptual side. Our system opens the way to discriminate between the two dominant contributions assigned to the filler in reinforcement mechanisms: the role of the particle interface and the one of the filler network. The first one includes interaction between the particles and the polymer, and the consequences on the chains dynamic[13-17]; the second one interconnected networks and/or aggregation[18]. Experimentally, it appears to be very difficult to separate the two effects, as one usually influences the other. With the new system, we are able to propose a simple alternative to act on the nanoparticles filler network without important modification of the interface. Indeed the simple orientation of the filler network does not involve modification of the matrix chains themselves (at variance with, for example, strain exerted on the full sample).

**Methods**

**Ferro-particles synthesis:**

The nanoparticles are made of maghemite (γ-$Fe_2O_3$). They are chemically synthesized in water by coprecipitation of $FeCl_2$ and $FeCl_3$ in an alkaline solution[19]. A size-sorting procedure[20] made at the end of the synthesis enables to reduce the polydispersity of the nanoparticles. The final particle size is obtained by magnetization measurements and by SANS experiments that give the same results. The size distribution is described by a log-normal law with a mean radius $R_0$ of 35 Å and a polydispersity index σ of 0.3. At the end of the synthesis the naked nanoparticles are stabilized at pH 12 by $N(CH_3)_4^+$ counterions. In order to increase their stability range (their PZC is 7.2 and they are only stable for pH > 10), their surface is coated by a thin shell of silica. The silica-coating process is based on a two-step procedure method described by Philipse *et al* [21] to obtain silica particles with a magnetic core, that we limit here to the first step. A solution of $SiO_2$, $Na_2O$ is added to a suspension of naked nanoparticles at pH 12 inducing the formation of silica oligomers in the solution. The oligomers adsorb on the nanoparticles surface. The concentrations of particles and silica correspond to a molar ratio [Fe]/[Si] = 1 and are chosen weak to prevent the formation of silica bridges between particles and thus irreversible aggregation. In order to eliminate silica oligomers in excess the suspension is dialyzed against a reservoir constituted by a NaOH solution at pH 10. The nanoparticles have the same size than before the silica-coating process (SANS measurements) but they are stable from pH 4 to pH 12. A very thin layer of silica has thus been coated on the particle that modifies their surface properties but not their magnetic properties.

**Composite nano-latex:**

The nanolatex was kindly provided by Rhodia. It is a solution of core-shell latex of poly(methyl methalycrate) (PMMA) and poly(butylacrylate) (PBuA), with a hydrophilic shell containing methacrylic acid that enables colloidal stability in water. The approximate size given by Rhodia is R ~ 200 Å. Before mixing it is dialyzed against the same reservoir as the one used for the dialysis of the silica-coated magnetic nano-particle to impose the same pH and ionic strength to the two components of the mixture.

The two solutions are then mixed with $\Phi_{nanoparticles}/\Phi_{nanolatex}$ = 0.05. The final mixture is degassed under primary vacuum about 1 day to avoid bubble formation. The solvent is then slowly evaporated during 4 days at T = 60°C under atmospheric pressure to get a homogeneous bubble-free film. During the drying step a magnetic field of 150 Gauss is imposed to the sample in the plane of the final film.

**Stress-strain isotherms and the reinforcement factor:**

Samples for stress-strain isotherms are brought to constant thickness using sandpaper. They are stretched up to rupture in a controlled constant-rate deformation ($\gamma$ = 0.0016 s$^{-1}$), at T = 60°C, i.e. well above the glass transition temperature of the matrix (nanolatex Tg = 33°C). The force F($\lambda$), where $\lambda$ = L/L$_0$ defines the elongation with respect to the initial length L$_0$, is measured with a HBM Q11 force transducer, and converted to (real) stress inside the material $\sigma$. The deformation of the film is assumed homogeneous, and incompressible. It is recalled that in our silica-latex system, the physico-chemical parameters that control the morphology of aggregates of silica nanoparticles are the silica concentration, the pH and the salinity in solution before solvent evaporation. We have shown previously that the rheological properties of the pure nano-latex matrix are pH-dependent[10]. This is one of the reasons why we analyze our data in terms of the nano-composite reinforcement factor R ($\lambda$) = $\sigma(\lambda)/\sigma_{latex}(\lambda)$, where the stress of the pure matrix $\sigma_{latex}(\lambda)$ has been measured at the same pH.

**Small Angle Neutron Scattering:**

Experiments have been performed at LLB (Saclay, France) on beam line PAXY, at ILL (Grenoble, France) on beam line D11 and at HMI (Berlin, Germany) on beam line V4. On PAXY, a set of two sample-to-detector distances 1 m and 5 m at wavelength 6Å give us a q range between $8.10^{-3}$ et 0.3 Å$^{-1}$. On D11, a set of two wavelengths, 6 and 12 Å, with four sample-to-detector distances, 1m, 5m 20m and 34m, gives us access to a q range comprised between $8.10^{-4}$ and 0.3 Å$^{-1}$. On V4, a set of three sample-to-detector distances 1m, 4m and 12m and two wavelengths, 6 and 12 Å, gives us a q range from $2.10^{-3}$ to 0.1 Å$^{-1}$. Data treatments have been done following standard procedures, with $H_2O$ as calibration standard, to get absolute intensities. Incoherent scattering and background contributions of the nano-latex films have been subtracted from the pure latex matrix scattering (the incoherent scattering from maghemite nanoparticles is null). In the simplest case of centro-symmetrical mono-disperse particles, the scattered intensity I(q) can be written as function of the wave vector q like I(q)=$\Phi$ $\Delta\rho^2$ V P(q) S(q), $\Phi$ is the volume fraction of the filler, $\Delta\rho$ the difference of scattering length density of particles with the one of the matrix or the solvent, V the volume of the particle, P(q) the form factor and S(q) the structure factor. For dilute systems, the structure factor is close to 1 because interactions are negligible. For concentrated systems with repulsive interactions, S (q) presents a correlation peak at q*=$2\pi$/D where D is the average distance between the scattering mass centres. For attractive systems, aggregates of initial particles can be formed with different compactness. For open systems, one gets a fractal organization of particles with a fractal dimension $D_f$ and the scattering spectra decay like $q^{-D_f}$ whereas for dense systems high compactness can be obtained due to close packing volume fraction.

Acknowledgements

We acknowledge the important help of Mireille Nuyadge (undergraduate trainee at LLB & LI2C) and Delphine Talbot (LI2C) during synthesis of ferrofluids (LI2C), and of Jean Chevalet (LI2C) for design of the film casting device under field.

**Figure 1** Sketch illustrating the route for incorporating magnetic nanoparticles in polymer films by latex film formation and the orientation of the fillers with a constant field.

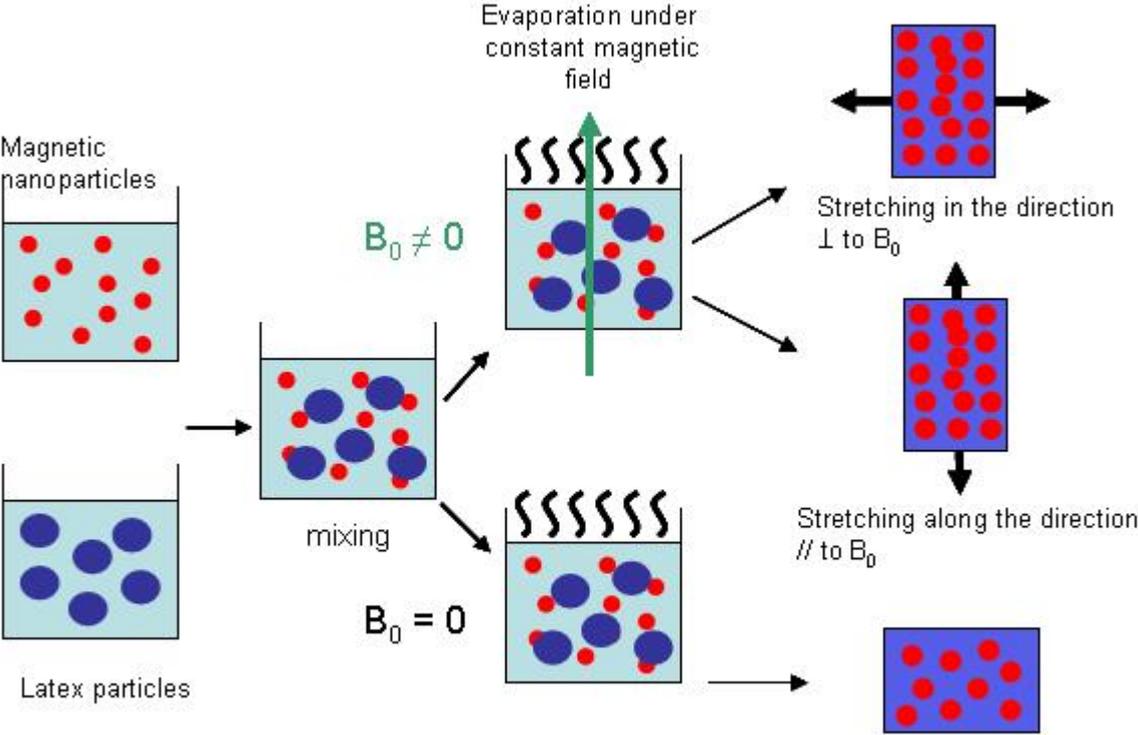

**Figure 2 a,** Stress-strain isotherm for: pure latex matrix, zero-field composite sample containing 5% in volume fraction of magnetic nanoparticles, and nano-latex films cast under constant field and stretched in directions parallel and perpendicular to the field. **b,** Reinforcement factor as function of the deformation for the field-cast samples under constant field. **c,** parallel on perpendicular ratio of reinforcements for the field-cast samples under constant field.

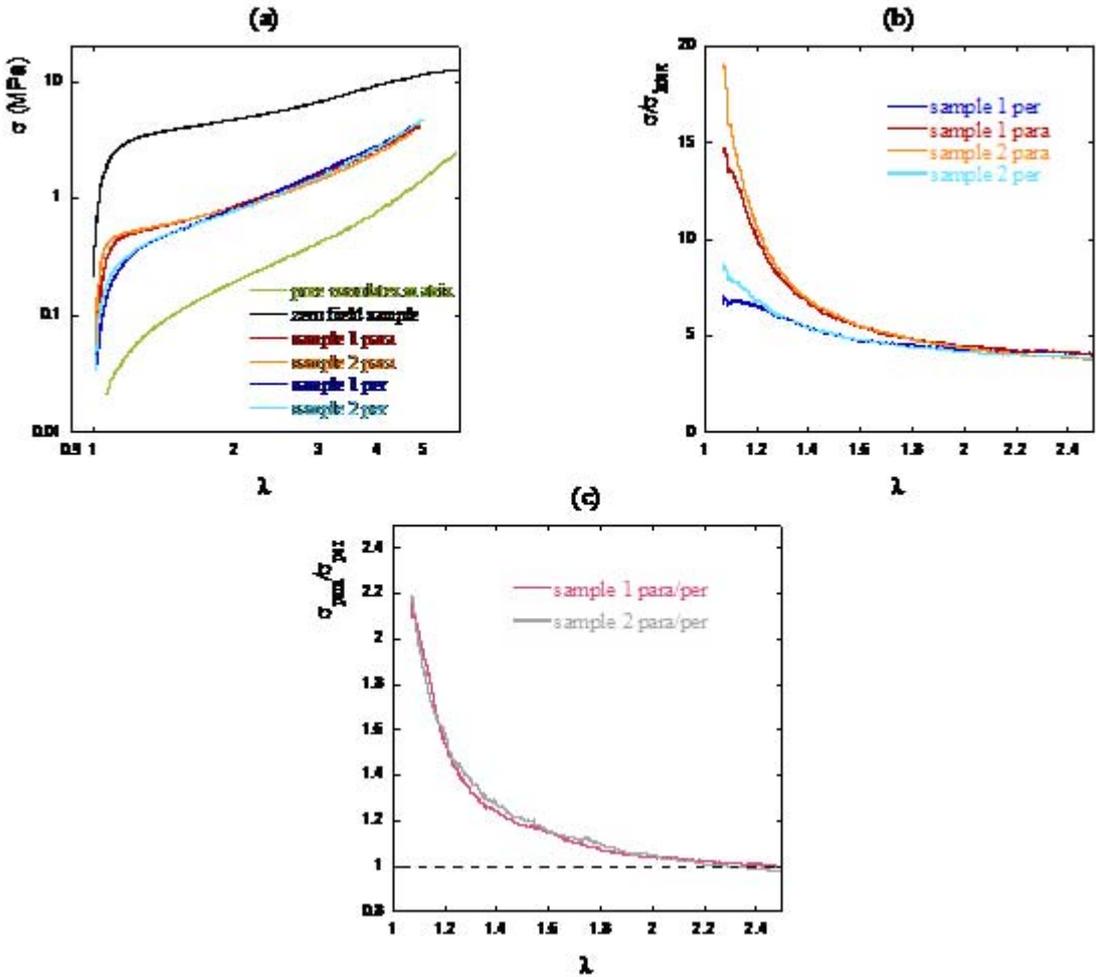

**Figure 3** SANS scattering of the samples at rest. From up to down: 2D spectra at the lowest q-range probed in the experiment DONNER DISTANCE ET LONGUEUR D ONDE ?; 1D scattered intensity; corresponding scheme of the structure. **a,** the initial aggregates of magnetic particles covered with silica layer, the full red line to the form factor of an isolated nanoparticle and the full blue line correspond to a cluster of 3-4 nanoparticles with $D^f = 1$ **b,** the zero-field composite sample and **c,** the nano-latex composite film evaporated under field.

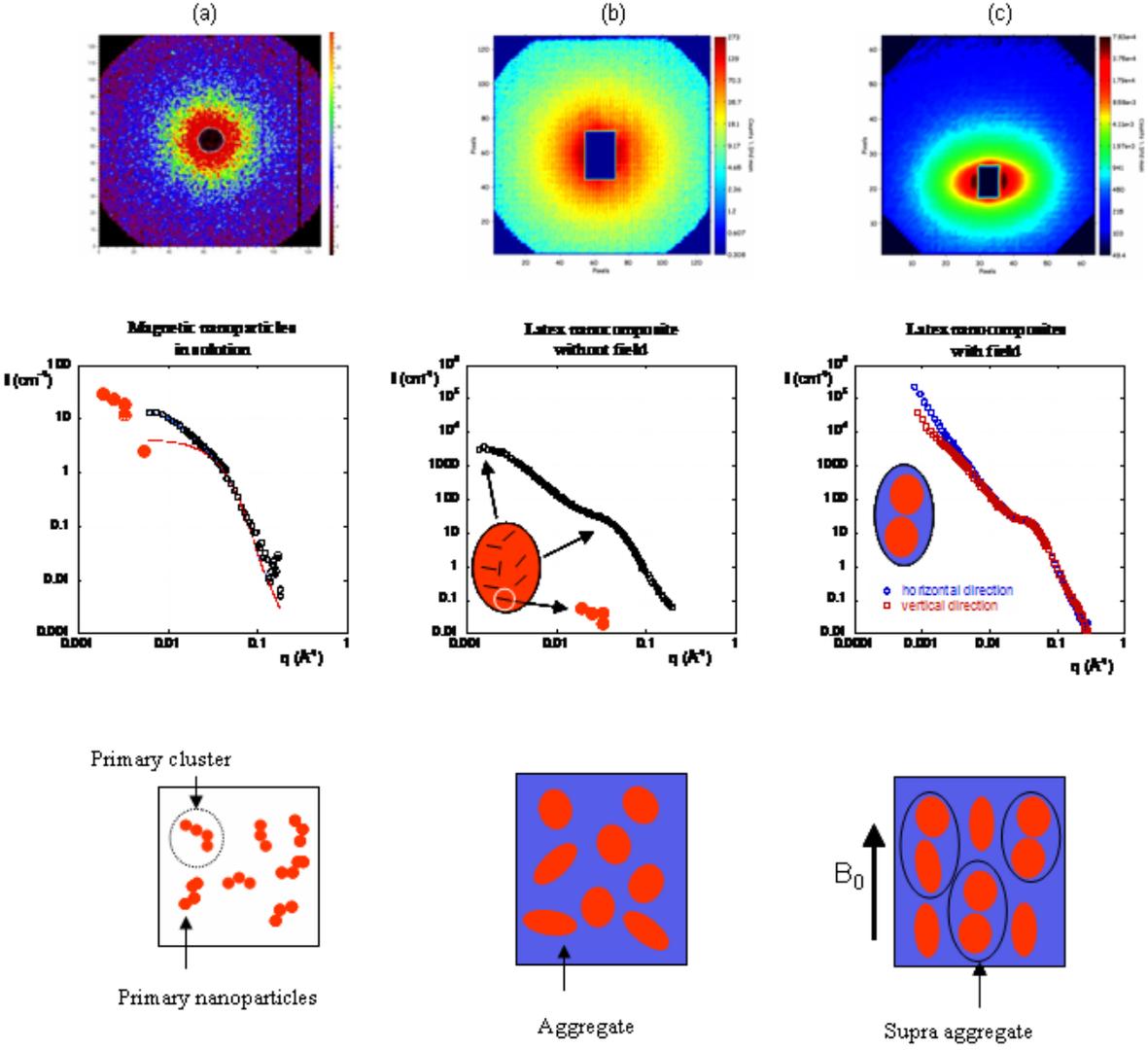

**Figure 4** 2D scattering of the stretched nano-latex film when the elongation direction is along the field-oriented fillers. Radial grouping of the S(q).q² versus q along the vertical **a,** and the horizontal direction of the picture **b,** for each elongation values and sketch of the structure in the real space. The hexagonal aspect of the 2D picture is due to the apparition of a correlation peak which corresponds to a characteristic distance between the aggregates along the stretching direction (black arrow for λ=1.5). The green arrow for λ=3 signals a smaller characteristic distance between the aggregates, which results from rearrangement by intercalation of the objects, under stretching Along the horizontal plan of the picture, the agglomerates are pushed closer.

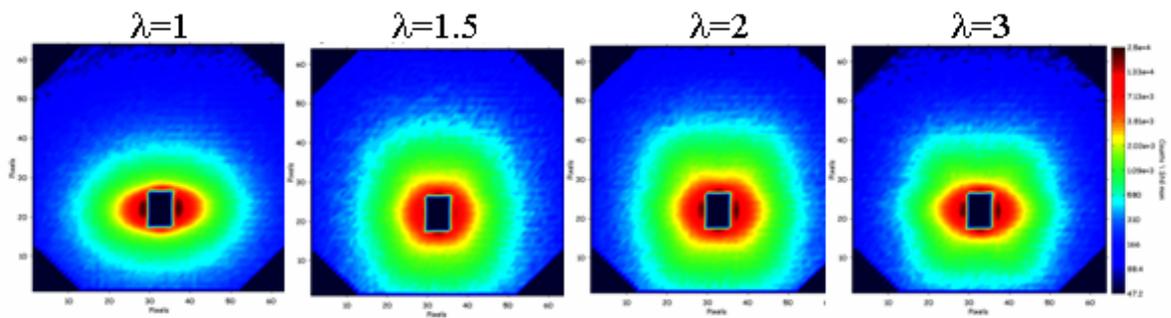

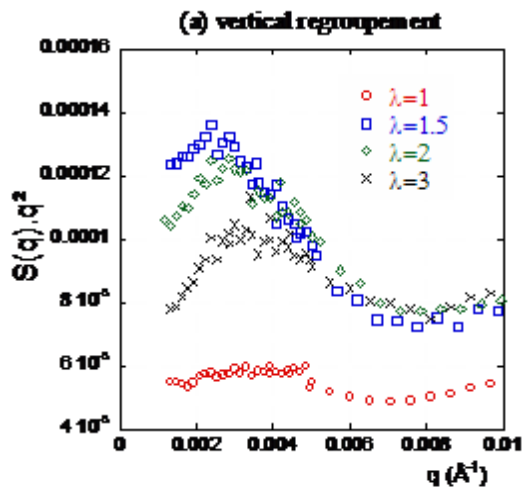

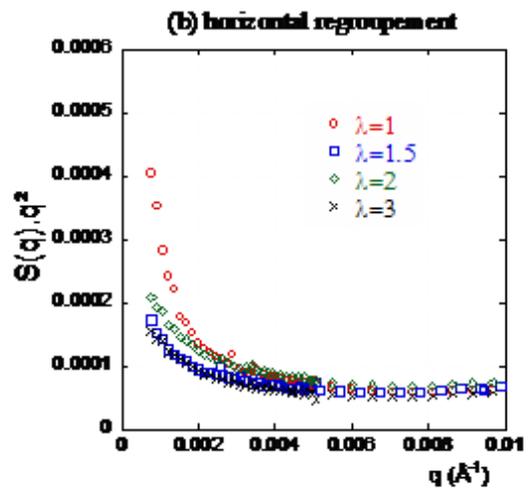

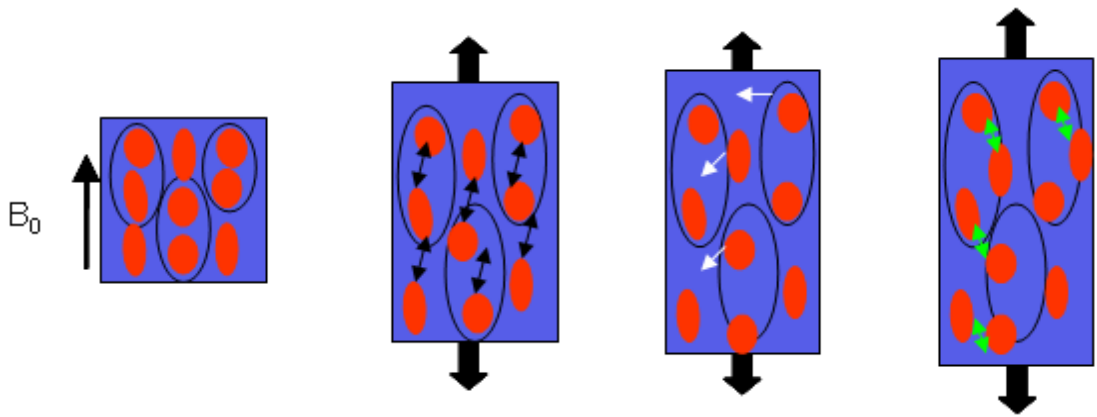

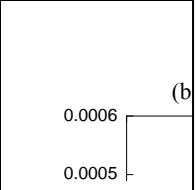

**Figure 5** SANS scattering of the stretched nano-latex film when the elongation direction is perpendicular to the field- oriented fillers. Radial intensities grouped along the vertical and the horizontal direction of the 2D picture as function of q. **a,** sample before stretching, **b,** elongated sample (λ=2), and zero-field cast isotropic sample (black dots).

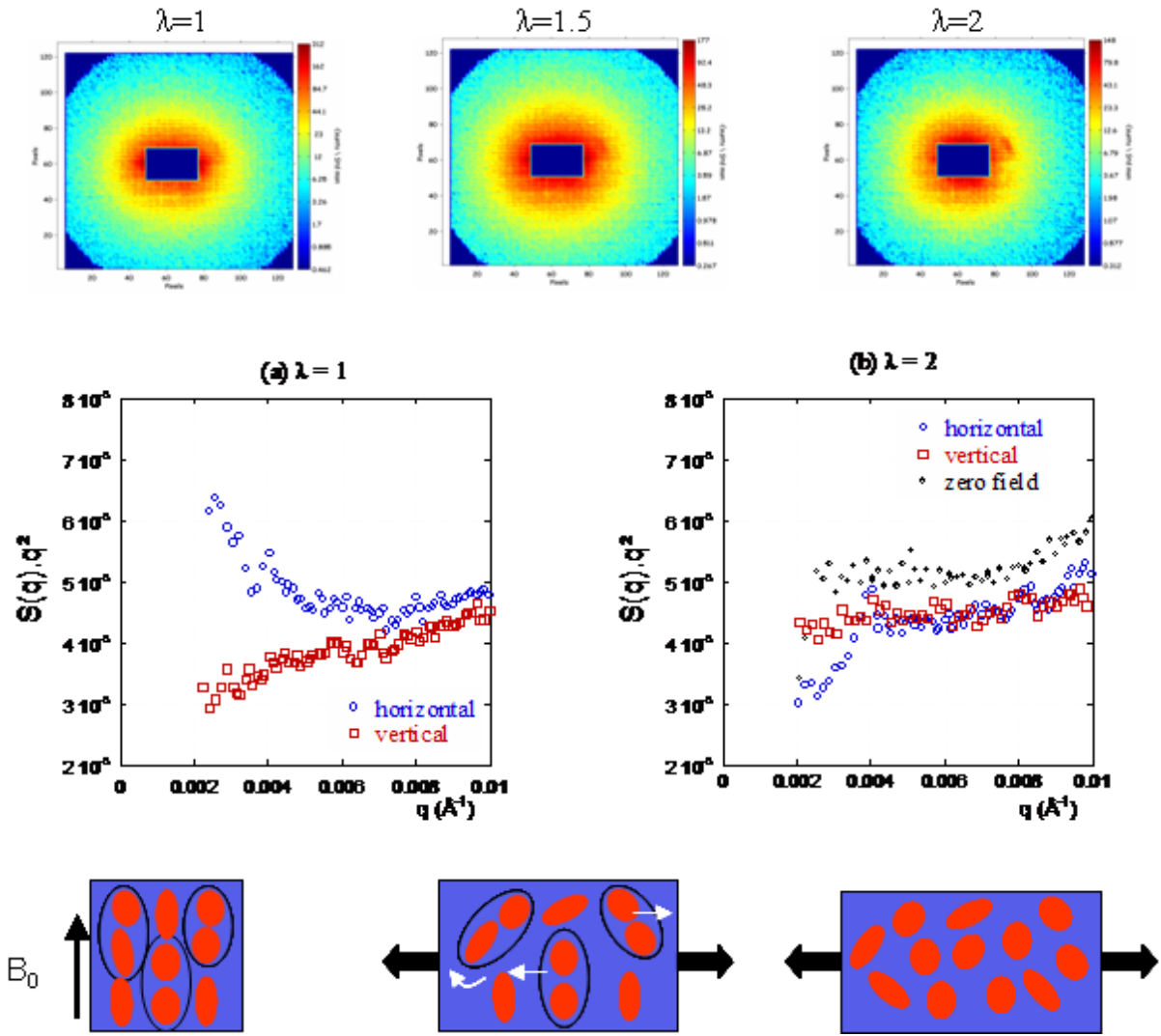